\documentclass[apjl,numberedappendix]{emulateapj}
\slugcomment{{\sc Accepted to ApJ:} November 20, 2016} 

\usepackage{multirow}
\usepackage[breaklinks,colorlinks,citecolor=blue]{hyperref}
\usepackage[all]{hypcap}

\usepackage{textcomp}

\newcommand\nobr{\mbox{-}}
\newcommand\ctht{\textit{c}\nobr C$_3$H$_2$}
\newcommand\moo{~\hbox{\textmu}m}
\begin{document}


\title{On the origin of C$_4$H and CH$_3$OH in protostellar envelopes\footnotemark[*]}\footnotetext[*]{Based on observations with the Kitt Peak 12 Meter telescope telescope and the Atacama Pathfinder EXperiment (APEX) telescope. The Kitt Peak 12 Meter telescope is operated by the Arizona Radio Observatory (ARO), Steward Observatory, University of Arizona. APEX is a collaboration between the Max Planck Institute for Radio Astronomy, the European Southern Observatory, and the Onsala Space Observatory.}


\author{
	Johan E. Lindberg\altaffilmark{1},
	Steven B. Charnley\altaffilmark{1},
	Martin A. Cordiner\altaffilmark{1,2}
}

\affil{$^1$NASA Goddard Space Flight Center, Astrochemistry Laboratory, Mail Code 691, 8800 Greenbelt Road, Greenbelt, MD 20771, USA, johan.lindberg@nasa.gov}
\affil{$^2$Department of Physics, Catholic University of America, Washington, DC 20064, USA}

\begin{abstract}

	The formation pathways of different types of organic molecules in protostellar envelopes and other regions of star formation are subjects of intense current interest. We here present observations of C$_4$H and CH$_3$OH, tracing two distinct groups of interstellar organic molecules, toward 16 protostars in the Ophiuchus and Corona Australis molecular clouds. Together with observations in the literature, we present C$_4$H and CH$_3$OH data from single-dish observations of 40 embedded protostars. We find no correlation between the C$_4$H and CH$_3$OH column densities in this large sample. Based on this lack of correlation, a difference in line profiles between C$_4$H and CH$_3$OH, and previous interferometric observations of similar sources, we propose that the emission from these two molecules is spatially separated, with the CH$_3$OH tracing gas that has been transiently heated to high ($\sim70$--100~K) temperatures, and the C$_4$H tracing the cooler large-scale envelope where CH$_4$ molecules have been liberated from ices. These results provide insight in the differentiation between \textit{hot corino} and \textit{warm carbon-chain chemistry} in embedded protostars.

\end{abstract}

\keywords{stars: formation --- ISM: molecules --- ISM: individual objects (Ophiuchus) --- astrochemistry}



\section{Introduction}\label{sec:intro}

\capstartfalse
\begin{deluxetable*}{lcclrlll}

	\tablecaption{Coordinates and other properties of the observed sources}

	\tablehead{\colhead{\textbf{Cloud}} & \colhead{R.A.} & \colhead{Dec.} & \colhead{$T_\mathrm{bol}$} & \colhead{$L_\mathrm{bol}$} &  \colhead{$\alpha_\mathrm{IR}$} & $N($H$_2$O$)_\mathrm{ice}$ & \colhead{Other common} \\ 
		\colhead{Source} & \colhead{(J2000.0)} & \colhead{(J2000.0)} & \colhead{[K]} & \colhead{[$L_\odot$]} & \colhead{} & [$10^{18}$~cm$^{-2}$] & \colhead{identifiers} }

	\startdata
	\textbf{$\rho$ Oph A} \\
	GSS 30 & 16:26:21.42 & $-24$:23:06.4 & $150$\tablenotemark{a} & $8.7\phantom{000}$ & $\phantom{0.}1.46$ & $1.41\pm0.20$\tablenotemark{b} & Elias 21 \\
	{[}GY92{]} 30 & 16:26:25.49 & $-24$:23:01.6 & $135$\tablenotemark{c} & $0.10\phantom{00}$ &  $\phantom{0.}0.87$ & ... \\
	ISO-Oph 21 & 16:26:17.23 & $-24$:23:45.4 & $490$\tablenotemark{a} & $0.083\phantom{0}$ & $\phantom{0.}0.69$ & ... & CRBR 12 \\
	J162614.6 & 16:26:14.62 & $-24$:25:08.4 & \phantom{0}... & ...\phantom{0000} & \phantom{0.}... & ... \\
	VLA 1623 & 16:26:26.42 & $-24$:24:30.0 & $\phantom{0}57$\tablenotemark{c} & $0.41\phantom{00}$ & $\phantom{0.}1.65$ & ... \\
	\textbf{$\rho$ Oph B} \\
	ISO-Oph 124 & 16:27:17.57 & $-24$:28:56.3 & $\phantom{0}68$\tablenotemark{a} & $2.6\phantom{000}$ & $\phantom{0.}0.25$ & ... \\
	J162728 & 16:27:28.45 & $-24$:27:21.0 & $310$\tablenotemark{a} & $0.48\phantom{00}$ & $-0.03$ & ... & IRS45, [GY92]~273 \\
	Oph-emb 5 & 16:27:21.83 & $-24$:27:27.6 & $180$\tablenotemark{a} & $0.019\phantom{0}$ & $-0.05$ & ... \\
	VSSG 17 & 16:27:30.18 & $-24$:27:43.4 & $530$\tablenotemark{a} & $0.93\phantom{00}$ & $-0.12$ & $1.70\pm0.25$\tablenotemark{d} & Elias 33, IRS47, [GY92]~279 \\
	\textbf{$\rho$ Oph C} \\
	WL 22 & 16:26:59.17 & $-24$:34:58.8 & $110$\tablenotemark{a} & $1.5\phantom{000}$ & $\phantom{0.}1.99$ & ... & ISO-Oph 90 \\
	\textbf{$\rho$ Oph F} \\
	ISO-Oph 137 & 16:27:24.61 & $-24$:41:03.4 & $191$\tablenotemark{c} & $0.33\phantom{00}$ & $\phantom{0.}1.01$ & $4.19\pm0.41$\tablenotemark{e} & CRBR85 \\
	\textbf{L1689N} \\
	IRAS 16293-2422 & 16:32:22.56 & $-24$:28:31.8 & $\phantom{0}47$\tablenotemark{a} & $16\phantom{.0000}$  & $\phantom{0.}5.03$ & ... \\
	\textbf{L1689S} \\
	ISO-Oph 203 & 16:31:52.45 & $-24$:55:36.2 & $240$\tablenotemark{a} & $0.13\phantom{00}$ & $\phantom{0.}1.07$ & ... \\
	\textbf{Solitary source} \\
	MMS126 & 16:28:21.61 & $-24$:36:23.4 & $\phantom{0}41$\tablenotemark{a} & $0.29\phantom{00}$ & $\phantom{0.}1.23$ & ... & IRAS 16253-2429\\
	\textbf{Corona Australis} \\
	CrA IRS1 & 19:01:50.68 & $-$36:58:09.7 & $373$\tablenotemark{f} & $12\phantom{.0000}$ & $\phantom{0.}0.92$\tablenotemark{g} & $2.45\pm0.24$\tablenotemark{e} & V710 \\
	CrA-44 & 19:02:58.67 & $-$37:07:35.9 & $148$\tablenotemark{f} & $1.3\phantom{000}$ & $\phantom{0.}1.66$\tablenotemark{g} & $5.26\pm1.88$\tablenotemark{e} & IRAS 32

	\enddata
	
	
	\tablenotetext{a}{\citet{evans09}}
	\tablenotetext{b}{A. Boogert (priv. comm.)}
	\tablenotetext{c}{\citet{enoch09}}
	\tablenotetext{d}{\citet{oberg11a}}
	\tablenotetext{e}{\citet{boogert08}}
	\tablenotetext{f}{\citet{chen97}}
	\tablenotetext{g}{\citet{peterson11}}
	
	\tablecomments{The references below for column 4 apply also to columns 5 and 6 if not otherwise indicated.}

	\label{tab:sourcelist}
\end{deluxetable*}
\capstarttrue

Organic molecules have been identified in many different environments in space. Typically, a distinction is made between oxygen-bearing molecules (believed to be formed on the surfaces of dust grains) and unsaturated hydrocarbon molecules, e.g.\ carbon chains \citep[believed to be formed predominantly in the gas phase; see e.g.][]{sakai13}. Examples of the former are methanol (CH$_3$OH), methyl formate, and dimethyl ether, which are typically found in the hot cores of high-mass star formation \citep{kurtz00}, but also in so-called \textit{hot corinos}, deeply embedded low-mass protostars with large abundances of complex organic molecules (COMs; where `complex' is defined as containing at least six atoms), thought to trace evaporation of the icy dust grain mantles caused by the high temperatures in the hot inner envelope \citep[e.g.][]{ceccarelli04}. The grain surface formation of such species may also be enhanced by the high radiation fields from the inner protostar \citep{oberg09}, but might, on the other hand, also be inhibited by desorption of key species such as CO \citep{lindberg15}. Lower abundances of COMs have more recently been detected toward more quiescent objects such as prestellar cores \citep{bacmann12} and protostellar cores \citep{oberg10}. The low-temperature desorption mechanism of CH$_3$OH remains unknown \citep{bertin16}. For at least some other COMs, a gas-phase formation scenario cannot be disregarded \citep{charnley97,taquet16}.

On the other hand, unsaturated carbon-chains such as cyanopolyynes (HC$_{2n+1}$N), polyyne radicals (C$_{2n}$H), and the related cyclic molecule \ctht\ have been detected
toward dark molecular clouds and carbon stars \citep{ohishi98}, but also more recently 
in large abundances toward the two deeply embedded low-mass protostars L1527 and IRAS~15398-3359 \citep{sakai09a,sakai09b}, here in combination with low abundances of oxygen-bearing COMs. \citet{sakai09a} proposed that this difference in the chemistry of deeply embedded protostars is a result of different collapse time scales. A slow collapse would let a large proportion of the carbon atoms to form CO before accumulation on the dust grains, which would lead to a \textit{hot corino} scenario, since the formation of oxygen-bearing organics is thought to be initiated by hydrogenation of CO in the ices to form H$_2$CO and CH$_3$OH. A rapid collapse would, however, make most carbon freeze out as C atoms, which will be hydrogenated to form CH$_4$ on the grain surfaces. Both CH$_3$OH and CH$_4$ are widely observe in interstellar ices \citep{boogert15}.
Desorption of CH$_4$ and C$_2$H$_2$ molecules can boost the abundances of many unsaturated hydrocarbons through gas-phase reactions \citep{markwick00,hassel08}. Thus, as a protostellar envelope heats up, at about 30~K CH$_4$ evaporation creates an environment with so-called \textit{warm carbon-chain chemistry} (WCCC; \citealt{sakai13}).

Single-dish and interferometric mapping observations have shown that oxygen-bearing COMs (e.g.\ CH$_3$OH) and unsaturated carbon chains (traced by HC$_3$N or C$_4$H) in the ISM typically are spatially separated; in at least nine protostellar sources (\citealt{pratap97}; \citealt{buckle06}; \citealt{cordiner12}; \citealt{lindberg12}).
Most of these sources show extended CH$_3$OH emission, thus the CH$_3$OH is not only tracing a central hot inner envelope, but also large-scale envelope emission. Individual pointings with single-dish telescopes cannot be used to study the spatial distribution of these species, but nevertheless show that they often have different excitation temperatures and linewidths \citep{lindberg15,lindberg16}, suggesting that they are spatially differentiated.

Observations also indicate that hot cores and \textit{hot corinos} can be chemically differentiated between O-rich and N-rich complex molecules \citep{herbst09} and there appears to be chemical differentiation also between \textit{hot corino} chemistry and WCCC \citep{sakai13}. In  particular, for WCCC sources, the abundances of unsaturated hydrocarbon molecules, such as C$_4$H, appear to be anti-correlated with abundances of CH$_3$OH and other COMs. However, HC$_3$N, and perhaps also HC$_5$N, may co-exist with CH$_3$OH in the hottest molecular cores, when the former are produced in a hot gas chemistry \citep{rodgers01}, suggesting that there may be relative differentiation between CH$_3$OH and specific classes of carbon-chain compounds. 
\citet{sakai13} have summarized the various chemical scenarios that could account for hydrocarbon/methanol anti-correlations through the CH$_4$/CH$_3$OH ice mantle abundance ratio.

\capstartfalse
\begin{deluxetable*}{lrrrrr}

	\tablecaption{Observed integrated intensities and rotational temperatures}

	\tablehead{\colhead{Source} & \colhead{C$_4$H} & \colhead{C$_4$H} & \colhead{CH$_3$OH} & \colhead{$T_\mathrm{rot}($\ctht$)$} & \colhead{$T_\mathrm{rot}($H$_2$CO$)$} \\ 
		\colhead{} & \colhead{85.634~GHz} & \colhead{85.673~GHz} & \colhead{218.440~GHz} & \colhead{} \\ 
		\colhead{} & \colhead{$E_\mathrm{u} = 20.5$~K} & \colhead{$E_\mathrm{u} = 20.6$~K} & \colhead{$E_\mathrm{u} = 45.5$~K} & \colhead{} \\ 
		\colhead{} & \colhead{[mK km s$^{-1}$]} & \colhead{[mK km s$^{-1}$]} & \colhead{[mK km s$^{-1}$]} & \colhead{[K]} & \colhead{[K]}}

	\startdata
	GSS 30 & $13\pm4$ & ...\phantom{0000} & $\phantom{00}102\pm19$ & $14.7\pm3.7$ & $35.4\pm1.2$ \\
	{[}GY92{]} 30 & $33\pm6$ & $25\pm6$ & $\phantom{00}725\pm18$ & $9.9\pm0.3$ & $36.2\pm0.4$ \\
	ISO-Oph 21 & $46\pm8$ & $26\pm6$ & $\phantom{00}175\pm29$ & ...\phantom{00000} & $30.4\pm1.4$ \\
	J162614.6 & $87\pm7$ & $41\pm5$ & $\phantom{00}484\pm28$ & ...\phantom{00000} & $33.6\pm0.6$ \\
	VLA 1623 & $36\pm5$ & $42\pm6$ & $\phantom{00}95\pm18$ & $11.2\pm0.5$ & $31.3\pm1.0$ \\
	ISO-Oph 124 & $26\pm6$ & $25\pm6$ & $\phantom{00}172\pm23$ & ...\phantom{00000} & $19.1\pm1.3$ \\
	J162728 & $71\pm7$ & $43\pm7$ & $\phantom{00}83\pm17$ & $9.0\pm0.2$ & $17.9\pm0.9$ \\
	Oph-emb 5 & $25\pm5$ & $36\pm5$ & $\phantom{00}417\pm23$ & $8.7\pm1.0$ & $15.5\pm0.7$ \\
	VSSG 17 & $78\pm7$ & $70\pm8$ & $\phantom{00}50\pm13$ & $10.9\pm0.4$ & $19.1\pm1.2$ \\
	WL 22 & $206\pm6$ & $188\pm6$ & $\phantom{00}62\pm16$ & $8.2\pm0.5$ & $<28.8$\phantom{00000} \\
	ISO-Oph 137 & $83\pm9$ & $76\pm8$ & $\phantom{00}52\pm17$ & $6.7\pm0.8$ & $<34.3$\phantom{00000} \\
	IRAS 16293-2422 (red) & $22\pm5$ & ...\phantom{0000} & $\phantom{00}2896\pm34$ & $14.4\pm0.3$ & $69.4\pm0.9$ \\
	IRAS 16293-2422 (blue) & $53\pm6$ & $42\pm7$ & $\phantom{00}189\pm21$ & $16.5\pm1.5$ & $123.8\pm3.1$ \\
	ISO-Oph 203 & $59\pm7$ & $35\pm7$ & $\phantom{00}76\pm12$ & $9.3\pm0.4$ & $22.6\pm1.3$ \\
	MMS126 & $38\pm5$ & $40\pm6$ & $\phantom{00}79\pm18$ & $8.6\pm0.3$ & $<22.4$\phantom{00000} \\
	CrA IRS1 & $19\pm6$ & ...\phantom{0000} & $\phantom{00}390\pm20$ & $8.9\pm1.0$ & $29.5\pm2.4$ \\
	CrA-44 & $42\pm9$ & $63\pm9$ & $<72\phantom{.0000}$ & $11.4\pm0.9$ & $<21.2$\phantom{00000} 
	\enddata
	
	
	
	\tablecomments{The CH$_3$OH integrated intensities and \ctht\ and H$_2$CO rotational temperatures were measured in APEX observations \citep{lindberg15,lindberg16}. In the sources where no \ctht\ rotational temperature could be calculated, the Ophiuchus average of 10.8~K was used. In the sources with only an upper limit on the H$_2$CO temperature, the upper limit was used to calculate a lower limit on the CH$_3$OH column density.}
	
	\label{tab:linestrengths}
\end{deluxetable*}
\capstarttrue

Very recently, \citet{graninger16} identified a tentative correlation between the gas-phase column densities of CH$_3$OH and C$_4$H in embedded protostars, and suggested that this is related to the simultaneous evaporation of CH$_4$ (which later forms C$_4$H) and CH$_3$OH from the dust grains. However, such a correlation would imply (1) that these two species with substantially different binding energies (1090~K for CH$_4$ and 5530~K for CH$_3$OH; \citealt{taquet16}) come off the grains simultaneously, and (2) that significant abundances of both CH$_3$OH and CH$_4$ co-exist in the ices. These conditions are in disagreement with the core temperatures of $\sim30$~K and different collapse time scales proposed by \citet{sakai09a}, as well as with maps of molecular clouds and protostellar envelopes which show that CH$_3$OH and carbon chains, presumably derived from CH$_4$ and C$_2$H$_2$ evaporation (e.g.\  C$_4$H and HC$_3$N), rarely spatially coexist \citep{buckle06,lindberg12}.
Unless an as yet unidentified non-thermal desorption mechanism acts to desorb CH$_3$OH in the cooler parts of the protostellar envelopes in our sample, thermal sublimation can be regarded as solely responsible for injecting CH$_3$OH and CH$_4$ into the gas. In cold sources, this thermal heating may be highly transient and initiated by grain-grain collisions perhaps leading to grain-mantle explosions releasing surface-layer molecules \citep{dhendecourt85,markwick00}. CH$_3$OH and H$_2$CO are both water-like species, and are as such thought to come off grains simultaneously despite their different binding energies \citep{viti04}, which is also in agreement with a strong correlation between the two species found in a large sample of embedded protostars \citep{lindberg16}.

To explore these issues further, we observed C$_4$H and CH$_3$OH toward 16 southern protostars and make comparisons with the sample of 15 northern sources investigated by \citet{graninger16}.

\section{Observations}

We observed 14 embedded protostars in the Ophiuchus molecular cloud and two embedded protostars in the Corona Australis molecular cloud using the Kitt Peak 12 Meter radio telescope in May 2016. A source list is presented in \autoref{tab:sourcelist}. The observations were performed with the Millimeter Autocorrelator (MAC) with a bandwidth of 300~MHz centered at 85.559~GHz and with a resolution of 48.8~kHz. This spectral setup covers the two $N=9$--$8$ spectral lines of C$_4$H at 85.634~GHz and 85.673~GHz.

The sample was selected from an unbiased APEX 218~GHz survey of the 38 embedded protostars in Ophiuchus \citep{lindberg16}. Only the 14 sources in which unsaturated hydrocarbon molecules (C$_2$D, \ctht, or HC$_3$N) were detected in our APEX survey were included in our sample in this search for C$_4$H. Two embedded protostars in the Corona Australis star-forming region with detections of these species \citep{lindberg15} were also included. In this work we also use CH$_3$OH observations from the APEX 218~GHz survey using the SHeFI APEX-1 receiver \citep{lindberg16}. 

In addition, we include results from \citet{graninger16}, who observed C$_4$H and CH$_3$OH toward 15 embedded protostars in the northern sky. We have also searched the literature and added all embedded protostars we could find where column densities of both C$_4$H and CH$_3$OH have been calculated from observations with single-dish telescopes (nine sources), thus making a total of 40 sources.

\section{Results}
\label{sec:results}

The measured integrated line intensities are presented in \autoref{tab:linestrengths} together with rotational temperatures of \ctht\ and H$_2$CO measured in APEX 218~GHz observations \citep{lindberg16}. Since the measured C$_4$H lines both have the same $E_\mathrm{u}$, we assumed that its temperature is equal to the \ctht\ rotational temperature to be able to estimate the column densities of C$_4$H in our sample. Likewise, the CH$_3$OH column density was calculated from the 218.440~GHz line intensity by adopting the H$_2$CO rotational temperature. The reason for the use of these different rotational temperatures is that the unsaturated hydrocarbons usually are spatially separated from species related to COMs, and often show different temperatures, LSR velocities, and line widths \citep[e.g.][]{lindberg12,lindberg15}. 
Furthermore, the CH$_3$OH rotational temperature often does not reflect the kinetic temperature due to sub-thermal excitation and other non-LTE effects \citep{bachiller98}, while the rotational temperature of the closely related molecule H$_2$CO is an excellent tracer of the kinetic temperature when using transitions of the same $J_\mathrm{u}$ \citep{mangum93}. We performed non-LTE calculations using RADEX \citep{vandertak07} and found that a reliable value on the CH$_3$OH column density can be computed with the rotational diagram method using the CH$_3$OH 218.440~GHz line intensity and the kinetic temperature at $n\sim10^6$~cm$^{-3}$ and $T\lesssim40$~K, in our case assuming $T_\mathrm{kin} = T_\mathrm{rot}($H$_2$CO$)$. However, inferring the rotational temperature from another species with possibly different excitation properties may introduce additional uncertainties on the computed C$_4$H and CH$_3$OH column densities. Such errors are difficult to quantify, and are not taken into account in our stated errors.

For some sources, only one spectral line of \ctht\ or H$_2$CO was detected in our APEX observations, and thus only upper limits on the rotational temperature could be measured by using upper limits of non-detected spectral lines. For the sources where $T_\mathrm{rot}($\ctht$)$ could not be measured, we used the average of this parameter in the Ophiuchus sources (10.8~K) instead of an upper limit since the range of \ctht\ rotational temperatures is very narrow. However, the H$_2$CO temperature can be influenced by external irradiation and often varies greatly throughout the cloud \citep{lindberg15}; thus, choosing an average value for the H$_2$CO temperature introduces considerable uncertainties. For these sources, we used the upper limit on $T_\mathrm{rot}($H$_2$CO$)$ to create lower limits for the CH$_3$OH column densities.
IRAS~16293-2422 shows two distinct velocity components of C$_4$H and CH$_3$OH, which we here treat separately, referred to as the red and blue components.

For consistency, we re-calculate the C$_4$H and CH$_3$OH column densities presented by \citet{graninger16} using our algorithm for rotational diagrams based on the method of \citet{goldsmith99}\footnote{Despite using the same line intensities, in our re-calculations, we obtained rotational temperatures a factor $\sim2$ times lower and column densities a factor $\sim5$ times lower than the values found by \citet{graninger16}.}. For the sources where a rotational temperature could not be computed, we used the sample average of 10.9~K for C$_4$H and 6.8~K for CH$_3$OH.
All column densities of C$_4$H and CH$_3$OH used here, both from this work and the literature, as well as all references, are presented in \autoref{tab:coldens}.

\begin{figure}[!tb]
	\epsscale{1.2}
	\plotone{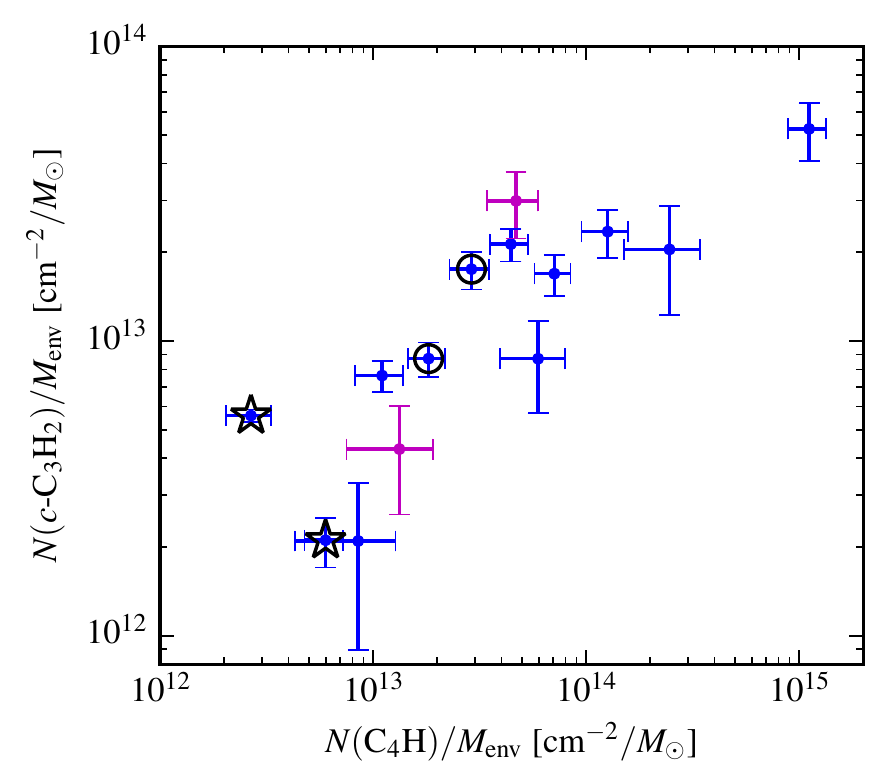}
	\plotone{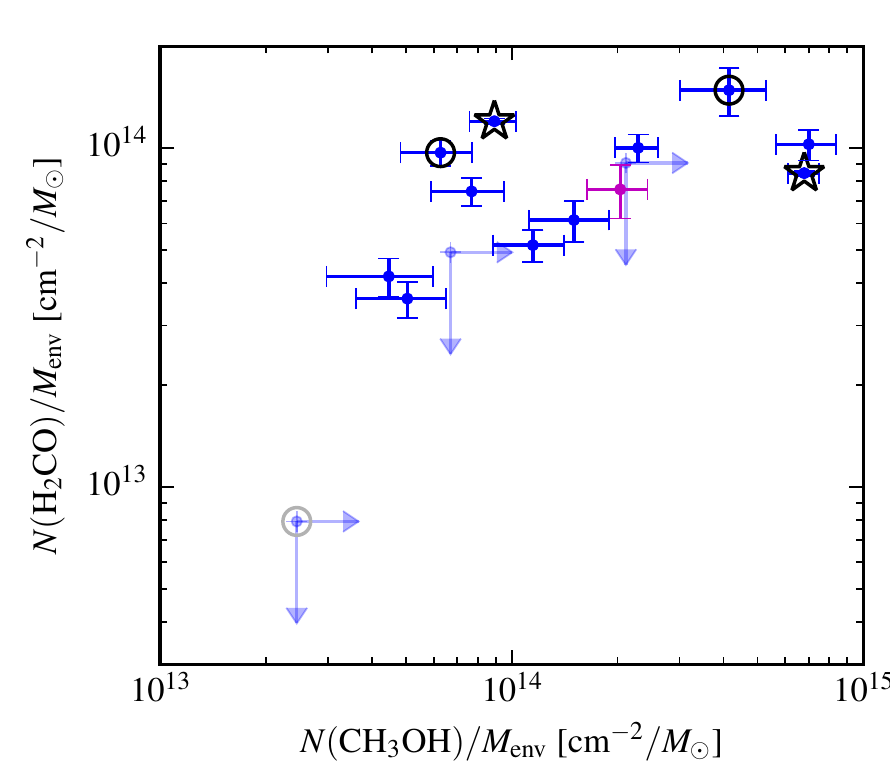}
	\caption{\textit{Top:} Correlation plot of C$_4$H and \ctht\ column densities normalised by envelope mass. \textit{Bottom:} Correlation plot of H$_2$CO and CH$_3$OH column densities normalised by envelope mass. See the caption of \autoref{fig:coldens} for legend.}
	\label{fig:fourmol}
\end{figure}

To justify the use of \ctht\ rotational temperatures for C$_4$H and H$_2$CO rotational temperatures for CH$_3$OH, we also investigate the correlation between the column densities of these molecules in \autoref{fig:fourmol}. To evaluate the possible correlation we use Spearman's rank correlation coefficient $\rho$, which tests the monotonicity of two sets of variables without requiring a linear relation. The \ctht\ shows a very strong correlation with C$_4$H ($\rho=0.82$, $p=0.0003$) and H$_2$CO is strongly correlated with CH$_3$OH ($\rho=0.62$, $p=0.033$). In a larger sample of embedded protostars, \citet{lindberg16} found an even stronger correlation between the column densities of H$_2$CO and CH$_3$OH. We find no correlation between H$_2$CO and C$_4$H (not plotted).

\capstartfalse
\begin{deluxetable*}{lrllrlllrl}

	\tablecaption{C$_4$H and CH$_3$OH column densities, LSR velocities, line widths, and envelope masses for all sources}

	\tablehead{\colhead{} & \colhead{} & \colhead{C$_4$H} & \colhead{} & \colhead{} & \colhead{CH$_3$OH} & \colhead{} & \colhead{} & \colhead{} \\
		\colhead{Source} & \colhead{$N$} & \colhead{$v_\mathrm{LSR}$} & \colhead{$\Delta v$} & \colhead{$N$} & \colhead{$v_\mathrm{LSR}$} & \colhead{$\Delta v$} & \colhead{Ref.} & \colhead{$M_\mathrm{env}$} & \colhead{Ref.} \\ 
		\colhead{} & \colhead{[$10^{13}$ cm$^{-2}$]} & \colhead{[km s$^{-1}$]} &  \colhead{[km s$^{-1}$]} & \colhead{[$10^{13}$ cm$^{-2}$]} & \colhead{[km s$^{-1}$]} & \colhead{[km s$^{-1}$]} & & \colhead{[$M_\odot$]}}

	\startdata
GSS 30 & $\phantom{0}0.19\pm0.09$ & $3.58\pm0.12$ & $0.64\pm0.16$ & $\phantom{00}1.7\phantom{0}\pm\phantom{0}0.4\phantom{0}$ & $3.08\pm0.12$ & $1.13\pm0.29$ & (1) & 0.22\phantom{0} & (8) \\
{[}GY92{]} 30 & $\phantom{0}0.58\pm0.14$ & $3.37\pm0.10$ & $0.89\pm0.27$ & $\phantom{0}12\phantom{.00}\pm\phantom{0}1\phantom{.00}$ & $3.28\pm0.01$ & $1.11\pm0.04$ & (1) & 0.53\phantom{0} & (8) \\
ISO-Oph 21 & $\phantom{0}0.66\pm0.38$ & $3.92\pm0.07$ & $0.90\pm0.15$ & $\phantom{00}2.8\phantom{0}\pm\phantom{0}0.6\phantom{0}$ & $3.14\pm0.18$ & $2.23\pm0.36$ & (1) & 0.24\phantom{0} & (9) \\
J162614.6 & $\phantom{0}1.1\phantom{0}\pm0.6\phantom{0}$ & $3.76\pm0.06$ & $1.31\pm0.13$ & $\phantom{00}7.9\phantom{0}\pm\phantom{0}0.9\phantom{0}$ & $2.76\pm0.05$ & $0.97\pm0.10$ & (1) & ...\phantom{000} & ... \\
VLA 1623 & $\phantom{0}0.69\pm0.13$ & $3.61\pm0.04$ & $0.70\pm0.09$ & $\phantom{00}1.5\phantom{0}\pm\phantom{0}0.3\phantom{0}$ & $3.69\pm0.04$ & $0.44\pm0.15$ & (1) & 0.24\phantom{0} & (8) \\
ISO-Oph 124 & $\phantom{0}0.48\pm0.28$ & $4.95\pm0.14$ & $1.08\pm0.27$ & $\phantom{00}2.9\phantom{0}\pm\phantom{0}0.7\phantom{0}$ & $3.53\pm0.12$ & $1.65\pm0.27$ & (1) & 0.070 & (8) \\
J162728 & $\phantom{0}1.3\phantom{0}\pm0.2\phantom{0}$ & $4.01\pm0.08$ & $1.26\pm0.24$ & $\phantom{00}1.5\phantom{0}\pm\phantom{0}0.4\phantom{0}$ & $4.22\pm0.06$ & $0.61\pm0.14$ & (1) & 0.29\phantom{0} & (8) \\
Oph-emb 5 & $\phantom{0}0.71\pm0.23$ & $4.02\pm0.06$ & $0.81\pm0.13$ & $\phantom{00}8.4\phantom{0}\pm\phantom{0}1.5\phantom{0}$ & $3.78\pm0.03$ & $0.97\pm0.09$ & (1) & 0.12\phantom{0} & (8) \\
VSSG 17 & $\phantom{0}1.3\phantom{0}\pm0.2\phantom{0}$ & $3.99\pm0.07$ & $1.34\pm0.18$ & $\phantom{00}0.85\pm\phantom{0}0.27$ & $3.19\pm0.05$ & $0.42\pm0.09$ & (1) & 0.19\phantom{0} & (8) \\
WL 22 & $\phantom{0}5.0\phantom{0}\pm0.9\phantom{0}$ & $3.83\pm0.00$ & $0.47\pm0.01$ & $>0.95\phantom{0000000}$ & $3.86\pm0.07$ & $0.66\pm0.13$ & (1) & 0.045 & (9) \\
ISO-Oph 137 & $\phantom{0}3.0\phantom{0}\pm1.1\phantom{0}$ & $4.03\pm0.03$ & $0.66\pm0.11$ & $>0.80\phantom{0000000}$ & $4.65\pm0.04$ & $0.23\pm0.09$ & (1) & 0.12\phantom{0} & (8) \\
IRAS 16293-2422\ (red) & $\phantom{0}0.31\pm0.07$ & $4.51\pm0.02$ & $0.29\pm0.04$ & $\phantom{0}79\phantom{.00}\pm\phantom{0}8\phantom{.00}$ & $4.24\pm0.01$ & $3.42\pm0.07$ & (1) & 1.2\phantom{00} & (8) \\
IRAS 16293-2422\ (blue) & $\phantom{0}0.69\pm0.14$ & $3.69\pm0.02$ & $0.42\pm0.04$ & $\phantom{0}10\phantom{.00}\pm\phantom{0}2\phantom{.00}$ & $0.46\pm0.20$ & $2.91\pm0.59$ & (1) & 1.2\phantom{00} & (8) \\
ISO-Oph 203 & $\phantom{0}1.0\phantom{0}\pm0.2\phantom{0}$ & $4.54\pm0.05$ & $0.71\pm0.10$ & $\phantom{00}1.2\phantom{0}\pm\phantom{0}0.3\phantom{0}$ & $4.78\pm0.07$ & $0.60\pm0.15$ & (1) & 0.080 & (8) \\
MMS126 & $\phantom{0}0.93\pm0.17$ & $4.13\pm0.03$ & $0.53\pm0.06$ & $>1.2\phantom{0}\phantom{0000000}$ & $3.58\pm0.14$ & $0.95\pm0.23$ & (1) & 0.51\phantom{0} & (8) \\
CrA IRS1 & $\phantom{0}0.40\pm0.17$ & $5.51\pm0.10$ & $0.71\pm0.26$ & $\phantom{00}6.1\phantom{0}\pm\phantom{0}1.0\phantom{0}$ & $5.88\pm0.03$ & $1.23\pm0.08$ & (1) & 0.30\phantom{0} & (10) \\
CrA-44 & $\phantom{0}0.94\pm0.23$ & $5.68\pm0.06$ & $0.77\pm0.13$ & ...\phantom{00000000} & $...$ & $...$ & (1) & 0.20\phantom{0} & (10) \\
B1-a & $\phantom{0}2.4\phantom{0}\pm1.2\phantom{0}$ & ... & ... & $\phantom{00}9.3\phantom{0}\pm\phantom{0}1.1\phantom{0}$ & ... & ... & (2) & 1.1\phantom{00} & (8) \\
SVS 4-5 & $\phantom{0}2.1\phantom{0}\pm0.8\phantom{0}$ & ... & ... & $\phantom{0}11\phantom{.00}\pm\phantom{0}1\phantom{.00}$ & ... & ... & (2) & 2.4\phantom{00} & (8) \\
B1-c & $\phantom{0}2.6\phantom{0}\pm1.1\phantom{0}$ & ... & ... & $\phantom{00}1.7\phantom{0}\pm\phantom{0}0.2\phantom{0}$ & ... & ... & (2) & 3.8\phantom{00} & (8) \\
IRAS 23238+7401 & $\phantom{0}1.4\phantom{0}\pm0.6\phantom{0}$ & ... & ... & $\phantom{00}1.8\phantom{0}\pm\phantom{0}0.3\phantom{0}$ & ... & ... & (2) & 0.28\phantom{0} & (11) \\
L1455 IRS3 & $\phantom{0}1.1\phantom{0}\pm0.5\phantom{0}$ & ... & ... & $\phantom{00}1.1\phantom{0}\pm\phantom{0}0.2\phantom{0}$ & ... & ... & (2) & 0.17\phantom{0} & (8) \\
B5 IRS1 & $\phantom{0}3.7\phantom{0}\pm1.7\phantom{0}$ & ... & ... & $\phantom{00}2.5\phantom{0}\pm\phantom{0}0.6\phantom{0}$ & ... & ... & (2) & 1.7\phantom{00} & (8) \\
L1455 SMM1 & $\phantom{0}1.8\phantom{0}\pm0.8\phantom{0}$ & ... & ... & $\phantom{00}1.4\phantom{0}\pm\phantom{0}0.2\phantom{0}$ & ... & ... & (2) & 0.50\phantom{0} & (8) \\
IRAS 03245+3002 & $\phantom{0}1.3\phantom{0}\pm0.7\phantom{0}$ & ... & ... & $\phantom{00}1.4\phantom{0}\pm\phantom{0}0.2\phantom{0}$ & ... & ... & (2) & 0.51\phantom{0} & (8) \\
L1014 IRS & $\phantom{0}3.1\phantom{0}\pm1.7\phantom{0}$ & ... & ... & $\phantom{00}0.85\pm\phantom{0}0.16$ & ... & ... & (2) & ...\phantom{000} & ... \\
IRAS 04108+2803 & $\phantom{0}0.55\pm0.25$ & ... & ... & $\phantom{00}1.8\phantom{0}\pm\phantom{0}0.5\phantom{0}$ & ... & ... & (2) & 0.080 & (12) \\
IRAS 03235+3004 & $\phantom{0}3.6\phantom{0}\pm1.2\phantom{0}$ & ... & ... & $\phantom{00}1.3\phantom{0}\pm\phantom{0}0.7\phantom{0}$ & ... & ... & (2) & 0.50\phantom{0} & (8) \\
L1489 IRS & $<0.93\phantom{000000}$ & ... & ... & $\phantom{00}0.53\pm\phantom{0}0.23$ & ... & ... & (2) & 0.10\phantom{0} & (13) \\
HH 300 & $\phantom{0}0.95\pm0.42$ & ... & ... & $\phantom{00}0.21\pm\phantom{0}0.09$ & ... & ... & (2) & 0.030 & (14) \\
IRAS 03271+3013 & $\phantom{0}1.4\phantom{0}\pm0.8\phantom{0}$ & ... & ... & $\phantom{00}0.35\pm\phantom{0}0.15$ & ... & ... & (2) & 0.46\phantom{0} & (8) \\
L1448 IRS1 & $<0.68\phantom{000000}$ & ... & ... & $\phantom{00}0.19\pm\phantom{0}0.09$ & ... & ... & (2) & ...\phantom{000} & ... \\
L1448 N & $\phantom{0}7.3\phantom{0}\pm1.7\phantom{0}$ & $4.83\pm0.06$ & $1.07\pm0.14$ & $\phantom{00}9.4\phantom{0}\pm\phantom{0}3.6\phantom{0}$ & ... & 1.2 & (3,4) & 0.15\phantom{0} & (15) \\
L1448 MM & $\phantom{0}4.7\phantom{0}\pm1.6\phantom{0}$ & $4.78\pm0.07$ & $0.98\pm0.17$ & $\phantom{00}8.2\phantom{0}\pm\phantom{0}2.2\phantom{0}$ & ... & 1.2 & (3,4) & 0.93\phantom{0} & (11) \\
NGC 1333 IRAS 2A & $\phantom{0}3.1\phantom{0}\pm1.3\phantom{0}$ & $7.83\pm0.07$ & $0.60\pm0.15$ & $\phantom{0}95\phantom{.00}\pm45\phantom{.00}$ & ... & 2.5 & (3,4) & 1.7\phantom{00} & (11) \\
NGC 1333 IRAS 4B & $<1.9\phantom{0}\phantom{000000}$ & ... & ... & $\phantom{0}74\phantom{.00}\pm17\phantom{.00}$ & ... & 1.2 & (3,4) & 2.0\phantom{00} & (11) \\
HH 211 & $\phantom{0}9.7\phantom{0}\pm1.1\phantom{0}$ & $9.08\pm0.02$ & $0.59\pm0.05$ & $\phantom{00}6.3\phantom{0}\pm\phantom{0}0.7\phantom{0}$ & 9.03 & 0.69 & (3,5) & 0.40\phantom{0} & (16) \\
L483 & $\phantom{0}4.2\phantom{0}\pm0.3\phantom{0}$ & $5.34\pm0.01$ & $0.52\pm0.03$ & $\phantom{00}2.6\phantom{0}\pm\phantom{0}1.0\phantom{0}$ & 5.29 & 0.43 & (3,5) & 1.1\phantom{00} & (11) \\
Serpens SMM4 & $\phantom{0}2.5\phantom{0}\pm0.6\phantom{0}$ & 8.08 & $1.40\pm0.15$ & $\phantom{0}16\phantom{.00}\pm\phantom{0}2\phantom{.00}$ & 8.07 & 1.96 & (3,5) & 5.3\phantom{00} & (17) \\
CrA IRS7B & $<0.70\phantom{000000}$ & ... & ... & $\phantom{0}16\phantom{.00}\pm\phantom{0}1\phantom{.00}$ & 5.8 & 2.1 & (3,6) & 2.2\phantom{00} & (18) \\
L1527  & $14\phantom{.00}\pm4\phantom{.00}$ & 5.84 & $0.40\pm0.01$ & $\phantom{00}6.3\phantom{0}\pm\phantom{0}1.0\phantom{0}$ & 5.8 & 0.39 & (3,7) & 0.91\phantom{0} & (11) 
	\enddata
	
	
	
	\tablecomments{All errors are $1\sigma$. LSR velocities and line widths are averages of the two C$_4$H lines when both are detected. References: (1) This work; (2) Recalculated values from \citet{graninger16} and \citet{oberg14}; (3) \citet{sakai09a}; (4) \citet{maret05}, re-calculated using H$_2$CO rotational temperatures calculated from line intensities in \citet{maret04}; (5) \citet{buckle02}; (6) \citet{lindberg15}; (7) \citet{sakai09b}; (8) \citet{enoch09}; (9) Calculated from submm fluxes and luminosities in \citet{jorgensen08}; (10) \citet{nutter05}; (11) \citet{jorgensen02}; (12) \citet{young03}; (13) \citet{brinch07}; (14) \citet{arce06}; (15) \citet{ciardi03}; (16) \citet{tanner11}; (17) \citet{hogerheijde99}; (18) \citet{lindberg12}.}
	
	\label{tab:coldens}
\end{deluxetable*}
\capstarttrue

\section{Discussion}

\begin{figure*}[!tb]
	\epsscale{1.2}
	\plotone{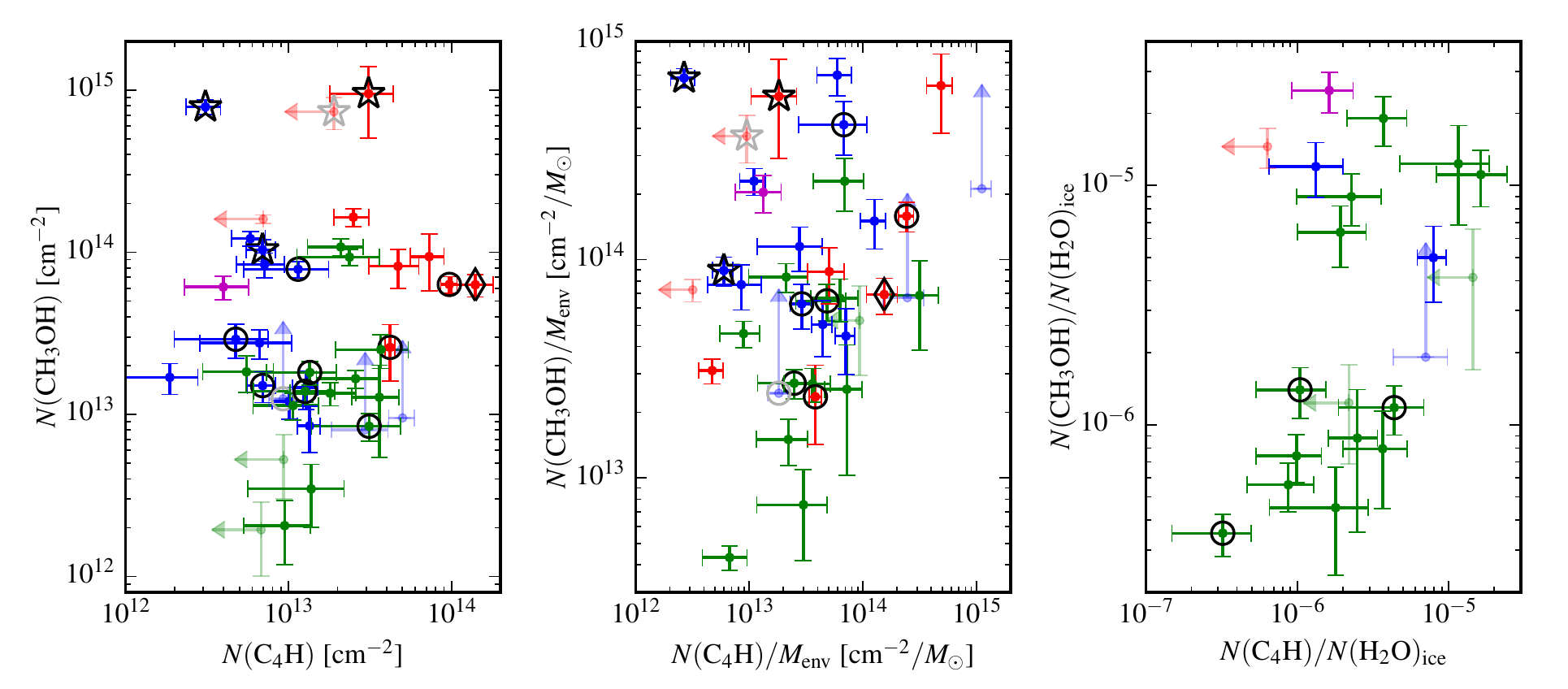}
	\caption{\textit{Left:} Measured C$_4$H and CH$_3$OH column densities toward embedded protostars. \textit{Middle:} The same column densities normalized by the envelope masses. \textit{Right:} The same column densities normalized by H$_2$O ice column densities. The data points represent sources in Ophiuchus (blue) and Corona Australis (magenta) from this work, sources from \citet{graninger16} (green), and other sources with data from the literature (red). \textit{Hot corino} sources are indicated with black stars, the WCCC source L1527 with a black diamond, and any other Class~0 sources (using the $T_\mathrm{bol} < 70$~K definition) with black rings. Upper/lower limit data points are semi-transparent.}
	\label{fig:coldens}
\end{figure*}

As discussed above, unsaturated hydrocarbons and COMs often show different excitation and line profile properties in studies of dark clouds and low-mass protostars. A comparison of the observed LSR velocities of Gaussian fits to the C$_4$H and CH$_3$OH lines in our sample shows that these agree within $3\sigma$ in only 4/16 sources (25\%) where both molecules have been detected, indicating that they often have different origins within the cloud (see \autoref{tab:coldens}). By contrast, the H$_2$CO and CH$_3$OH LSR velocities agree within $3\sigma$ in 12/16 cases (75\%), and the \ctht\ and C$_4$H velocities in 9/13 cases (69\%). \citet{graninger16} do not provide the observed LSR velocities for their sources, and we can thus not include them in this comparison. For the other literature sources, this information is available in some cases (we can compare line widths in 7 sources and LSR velocities in 4). In most cases they appear to be matching well, contrary to what we find in our sample.

The left panel of \autoref{fig:coldens} shows a correlation plot of the column densities of C$_4$H and CH$_3$OH from this work as well as the literature. A positive correlation between C$_4$H and CH$_3$OH was suggested by \citet{graninger16}, although such a trend was only significant if normalizing the column densities with H$_2$O ice column densities. Using recalculated column densities, such a trend is still plausible in that sample, but on the other hand, our data points from Ophiuchus suggest a negative correlation. None of these trends are significant. When combining these two datasets with the nine data points of single-dish measurements of C$_4$H and CH$_3$OH from the literature we find no discernible correlation.

To normalize the observed column densities, \citet{graninger16} use $N($H$_2$O$)_\mathrm{ice}$ as a proxy for the envelope material along the line of sight. However, C$_4$H forms from evaporated CH$_4$ in the gas phase \citep[e.g.][]{sakai09a}. Thus, we instead desire a tracer of the total amount of gas in the beam, for which the envelope mass ($M_\mathrm{env}$; typically derived from dust continuum observations at wavelengths $\sim1$~mm; e.g. \citealt{enoch09}) is a good proxy. The envelope mass is also much easier to measure than the H$_2$O ice column density, which requires strong flux at 3\moo, making such measurements towards Class~0 sources extremely difficult. The middle panel of \autoref{fig:coldens} shows the C$_4$H and CH$_3$OH column densities normalized by $M_\mathrm{env}$, and no correlation is detected. Neither do we find a correlation if the \textit{hot corino} sources or all Class~0 sources are disregarded. To account for the possibility that the CH$_3$OH actually coincides with the cool C$_4$H, we also calculated $N($CH$_3$OH$)$ assuming that $T_\mathrm{rot}($CH$_3$OH$) = T_\mathrm{rot}($\ctht$)$, and still find no correlation.

We also show the column densities normalized with $N($H$_2$O$)_\mathrm{ice}$ (\autoref{fig:coldens}, right panel), but since infrared observations of the molecular ice absorption are unavailable for most sources in our sample we can only add a few data points to the correlation plot of \citet{graninger16}. For these data points (excluding those that are upper limits) the correlation coefficient $\rho = 0.48$ with a $p$-value of $0.06$, indicating that no statistically significant correlation can be found. If only including the data points of the \citet{graninger16} sample, $\rho = 0.74$ with a $p$-value of 0.003, indicating a strong correlation. $N($H$_2$O$)_\mathrm{ice}$ measurements of the remaining sources in the sample is necessary to confirm or disprove this correlation, but considering the spread of the Ophiuchus data points (blue data points, left panel of \autoref{fig:coldens}) it is likely that also the $N($H$_2$O$)_\mathrm{ice}$-normalized correlation plot will show a larger spread when adding those data. Many of the missing sources are Class~0 objects, bound to be very faint at 3\moo, and measurements of $N($H$_2$O$)_\mathrm{ice}$ will prove to be difficult before the advent of the \textit{James Webb} Space Telescope (JWST).

The \textit{hot corino} sources (shown with black stars in \autoref{fig:coldens}) have significantly higher CH$_3$OH column densities than the sample average, which is expected. If treating the Class~0 sources in the sample separately (as defined by $T_\mathrm{bol}<70$~K) there is no discernible trend between the CH$_3$OH and C$_4$H column densities.

\section{Conclusions}

Recent observations led \citet{graninger16} to conclude that the gas-phase abundances of C$_4$H and CH$_3$OH in protostellar envelopes {\it are} in fact correlated. We have observed C$_4$H toward 16 protostars and, when these data are combined with a re-analysis of those of \citeauthor{graninger16} and others from the literature, we find no definitive evidence for such a correlation. The fact that CH$_3$OH is detected in these sources points to the dust and gas having been heated to $\sim 70$--$100$~K making the CH$_3$OH thermally sublimate. In the sources containing a protostar this may be due to, e.g., a previous short outburst (\citealt{audard14}; \citealt{taquet16}). Our observations support the view that CH$_3$OH and C$_4$H respectively reside in warmer and cooler regions of the protostellar environment, where carbon-chain formation is respectively suppressed and enhanced. The most dense and hot regions will have higher H$_2$O abundances, leading to destruction of C$^+$, making the reaction of C$^+$ with CH$_4$ (which leads to the formation of hydrocarbons) less efficient than in the lower-density, H$_2$O-free, outer envelope \citep{sakai13}. In the known WCCC sources, which have temperatures $\sim$ 30~K, the presence of gas-phase CH$_3$OH and HDO in the central envelope \citep{jorgensen13,bjerkeli16} indicates a recent accretion burst causing ice-mantle evaporation.

In the near future JWST will enable detections of ices in many faint sources representing different stages of the star formation process: from field stars tracing ices in cold dark clouds, to the youngest, most deeply embedded protostars. It will then be possible to explore in more detail expected correlations and anti-correlations between gas-phase and ice-phase chemistries.

\acknowledgments

We thank the anonymous referee whose detailed report helped improve the analysis.

This research was supported by an appointment to the NASA Postdoctoral Program at the NASA Goddard Space Flight Center to J.E.L., administered by Universities Space Research Association through a contract with NASA, and by NASA's Emerging Worlds Program.

We gratefully acknowledge the support from the staff at Arizona Radio Observatory and the APEX telescope.

\bibliographystyle{apj}
\bibliography{c4h_paper}






\end{document}